\documentclass[aps,prc,amsfonts,preprintnumbers,superscriptaddress,showpacs,nofootinbib]{revtex4}

\usepackage{graphics}
\usepackage{psfrag}
\usepackage{epsfig}
\usepackage{epsf}
\usepackage{float}

\newcommand{\beq}{\begin{equation}}
\newcommand{\eeq}{\end{equation}}
\newcommand{\beqa}{\begin{eqnarray}}
\newcommand{\eeqa}{\end{eqnarray}}
\newcommand{\nn}{\nonumber \\ }

\def\sl#1{\slash{\hspace{-0.2 truecm}#1}}

\begin{document}

\preprint{FZJ-IKP-TH-2008-24}
\preprint{HISKP-TH-08/24}

\title{On-Shell Consistency of the Rarita-Schwinger Field Formulation}

\author{H.~Krebs}
\email[]{Email: hkrebs@itkp.uni-bonn.de}
\affiliation{Helmholtz-Institut f\"ur Strahlen- und Kernphysik (Theorie)
and Bethe Center for Theoretical Physics,
 Universit\"at Bonn, D-53115 Bonn, Germany}
\author{E.~Epelbaum}\email{e.epelbaum@fz-juelich.de}
\affiliation{Forschungszentrum J\"ulich, Institut f\"ur Kernphysik (IKP-3) and
J\"ulich Center for Hadron Physics, \\
             D-52425 J\"ulich, Germany}
\affiliation{Helmholtz-Institut f\"ur Strahlen- und Kernphysik (Theorie)
and Bethe Center for Theoretical Physics,
 Universit\"at Bonn, D-53115 Bonn, Germany}
\author{Ulf-G. Mei{\ss}ner}\email{meissner@itkp.uni-bonn.de}
\affiliation{Helmholtz-Institut f\"ur Strahlen- und Kernphysik (Theorie)
and Bethe Center for Theoretical Physics,
 Universit\"at Bonn, D-53115 Bonn, Germany}
\affiliation{Forschungszentrum J\"ulich, Institut f\"ur Kernphysik (IKP-3) and
J\"ulich Center for Hadron Physics, \\
             D-52425 J\"ulich, Germany}
\date{\today}

\begin{abstract}
We prove that any bilinear coupling of a massive spin-$3/2$ field can be brought into 
a gauge invariant form suggested by Pascalutsa by means of a non-linear field redefinition. 
The corresponding field transformation is given explicitly in a closed form
and the implications for chiral effective field theory with explicit
$\Delta$(1232) 
isobar degrees of freedom are discussed.
%Assuming the consistency of the gauge invariant formulation we conclude that
%the well-known inconsistencies 
%of the Rarita-Schwinger formulation such as e.g.~acausal propagation can only show up  
%off mass shell. 
%On the mass-shell Lorentz-Covariance has to be restored.
\end{abstract}

\pacs{13.75.Cs,21.30.-x}

\maketitle

%\vspace{-0.2cm}

%%%%%%%%%%%%%%%%%%%%%%%%%%%%%%%%%%%%%%%%%%%%%%%%%%%%%%%%%%%%%%%%%%%%%%%%%%%%%%%%%
\section{Introduction}
\def\theequation{\arabic{section}.\arabic{equation}}
\setcounter{equation}{0}
\label{sec:intro}

A consistent formulation of a field theory involving interacting spin-$3/2$ degrees 
of freedom is an old and much studied problem, see
e.g.~Refs.~\cite{Johnson:1960vt,Velo:1969bt,Benmerrouche:1989uc}.
Recently, it has attracted renewed interest  
in the context of chiral effective field theory of QCD with explicit $\Delta$(1232) isobar degrees 
of freedom and its applications to e.g.~pion-nucleon scattering or
photo-nuclear processes, see Refs.~\cite{Pascalutsa:2006up,Bernard:2007zu} for recent review
articles. 

In the commonly used  Rarita-Schwinger formalism, the spin-$3/2$
field is represented by a vector-spinor 
$\psi_\mu$ (here and in what follows, we omit the spinor indices). It is well known 
how to write down the most general free Lagrangian for $\psi_\mu$ which describes the proper 
number of degrees of freedom \cite{Moldauer:1955}. The unphysical spin-$1/2$
degrees of freedom are projected out
in the resulting free equations of motion. It is much more difficult to ensure the decoupling of 
the unphysical degrees of freedom in the case of interacting  spin-$3/2$ fields. An elegant way to 
achieve this goal is to require that all interactions have the same type of gauge invariance as the 
kinetic term of the  spin-$3/2$ field \cite{Pascalutsa:1998pw}. This requirement of gauge invariance 
is, however, not compatible with the non-linear realization of the chiral symmetry by 
Coleman, Callan, Wess and Zumino \cite{Coleman:1969sm,Callan:1969sn}, which is
commonly adopted in chiral 
effective field theory and ensures the chiral invariance of the effective Lagrangian 
on a term-by-term basis.  
In this context, an important observation was made by Pascalutsa in 
Ref.~\cite{Pascalutsa:2000kd}, 
who showed that any gauge non-invariant linear coupling of the spin-$3/2$ fields
can be brought into the gauge invariant form via a suitably chosen field redefinition. 
In that work, a conjecture was made that such a field redefinition should also
be possible  for bilinear 
couplings but no explicit form for this transformation was given. In this work we fill this gap and 
and prove that such a transformation indeed exists for arbitrary bilinear couplings. Moreover, we 
are able to give the transformation explicitly in a closed form. 
%Our result
%not only encompasses but also transcends the findings of Ellis and
%Tang~\cite{Tang:1996sq}, who showed that the so-called off-shell parameters
%related to the spin-$1/2$ components are redundant as they can be absorbed in
%the low-energy constants accompanying certain local interaction operators.

Our paper is organized as follows. In section~\ref{sec1} we employ the functional integral 
technique to show that an interacting theory with gauge-invariant couplings of 
the spin-$3/2$ fields has the same constraints as the free theory. 
The non-linear field transformation which brings the   
gauge non-invariant bilinear couplings into the gauge invariant ones is 
discussed in section~\ref{sec2}. A brief discussion on the so-called off-shell
parameters can be found in section~\ref{sec:offshell}.   
In section~\ref{sec3} we show how our central result can be understood at the 
level of Feynman diagrams.  We end with a summary and conclusions.  
Some technicalities are relegated to the appendix.

%%%%%%%%%%%%%%%%%%%%%%%%%%%%%%%%%%%%%%%%%%%%%%%%%%%%%%%%%%%%%%%%%%%%%%%%%%%%%%%%%
\section{Spin-$3/2$ fields with gauge-invariant interactions}
\def\theequation{\arabic{section}.\arabic{equation}}
\setcounter{equation}{0}
\label{sec1}
The most general Lagrangian for spin-$3/2$ fields can be written in the
following form (after setting the point-transformation parameter $A=-1$) 
\cite{Moldauer:1955,Tang:1996sq}
\beq
{\cal
  L}=\bar{\psi}^\mu((i\sl{\partial})_{\mu\nu}-m_{\mu\nu}+V_{\mu\nu})\psi^\nu=
\bar{\psi}(i\sl{\partial}-m+V)\psi \,.  
\eeq
Here and in what follows, we use the short-hand notation:
\beq
(i\sl{\partial})^{\mu\nu}=\gamma^{\mu\nu\alpha}i\partial_\alpha,\quad
m^{\mu\nu}=\gamma^{\mu\nu}m,
\eeq  
with $m$ being the mass of the spin-$3/2$ field and $\gamma^\nu$ the Dirac
matrices. The quantities $\gamma^{\mu\nu}, \gamma^{\mu\nu\alpha}$ are
defined  according to 
\beq
\gamma^{\mu\nu}=\frac{1}{2}[\gamma^\mu,\gamma^\nu],\quad
\gamma^{\mu\nu\alpha}=\frac{1}{2}\{\gamma^{\mu\nu},\gamma^\alpha\}
\eeq
and are completely antisymmetric with respect to the Dirac indices. 
We will also need the inverse of the mass
operator $m^{\mu\nu}$ which in $d$ dimensions takes the form
\beq
\left[\frac{1}{m}\right]^{\mu\nu}=-\frac{1}{m}
\left( g^{\mu\nu} + \frac{1}{1-d}\,\gamma^{\mu}\gamma^{\nu}\right)~.
\eeq
Notice that because of the completely antisymmetric nature of 
$\gamma^{\mu\nu\alpha}$, the free massless Lagrangian is invariant under the 
gauge transformation
\beq\label{gaugetrafo}
\psi_\mu\rightarrow\psi_\mu+\partial_\mu\epsilon,
\eeq
where $\epsilon$ is an arbitrary spinor. We call the interacting field theory
gauge-invariant if all \emph{interaction} terms are invariant under this gauge
transformation.
%the corresponding massless interacting theory is invariant under this
%gauge transformation.

A gauge-invariant theory can be easily quantized using the standard path-integral 
technique. Indeed the path-integral formulation leads, by using the 
Faddeev-Popov trick, very naturally to the constraints which
have the same form as in the free field theory. Let us denote the action 
of a gauge-invariant theory by
\beq 
S=S[\bar{\psi}_\mu,\psi_\mu,\phi]=S_0[\bar{\psi}_\mu,\psi_\mu,\phi]-
\int d^4 x \, \bar{\psi}m\psi.
\eeq
Here we denote by $S_0$ the gauge invariant part of the theory
\beq
S_0[\bar{\psi}_\mu+\partial_\mu\bar{\epsilon},\psi_\mu+\partial_\mu\epsilon,\phi]=S_0[\bar{\psi}_\mu,
\psi_\mu,\phi]\,,  
\eeq
and  $\phi$ represents all other fields which interact with the spin-$3/2$ particle.
The partition function in the path-integral formulation is given by
\beq
Z=\int D\psi_\mu D\psi^\dagger_\mu D\phi \exp\left\{i S\right\}.
\eeq
Inserting the Faddeev-Popov $1$-operator
\beq
1=\int D\epsilon\,D\bar{\epsilon}\,
\delta((\bar{\psi}_\mu+\partial_\mu\bar{\epsilon})\gamma^\mu)\,
\delta(\gamma_\mu(\psi^\mu-\partial^\mu\epsilon))
\frac{1}{Det[-\gamma_\mu\partial^\mu]^2}
\eeq 
we get
\beq
Z=\int D\psi_\mu D\bar{\psi}_\mu D\phi\,D\epsilon\, D\bar{\epsilon}\,
\delta((\bar{\psi}_\mu+\partial_\mu\bar{\epsilon})\gamma^\mu)\,
\delta(\gamma_\mu(\psi^\mu-\partial^\mu\epsilon))
\frac{1}{Det[-\gamma_\mu\partial^\mu]^2} \exp\left\{i S\right\}~.
\eeq
Using gauge-invariance we can rewrite the action as
\beq
S[\bar{\psi}_\mu,\psi_\mu,\phi]=S[\bar{\psi}_\mu+\partial_\mu\bar{\epsilon},
\psi_\mu-\partial_\mu\,\epsilon,\phi]+\bar{\psi}_\mu\gamma^{\mu\nu}m\,
\partial_\nu\epsilon+\bar{\epsilon}\,\partial_\mu\gamma^{\mu\nu}m\,\psi_\nu.
\eeq
Substituting additionally the spin-$3/2$ fields by
\beq
\bar{\psi}_\mu\rightarrow\bar{\psi}_\mu-\partial_\mu\bar{\epsilon},\quad
\psi_\mu\rightarrow\psi_\mu+\partial_\mu\epsilon,
\eeq
we obtain for the partition function
\beq
Z=\frac{1}{Det[-\gamma_\mu\partial^\mu]^2}
\int D\psi_\mu D\bar{\psi}_\mu D\phi\,D\epsilon\, D\bar{\epsilon}\,
\delta(\bar{\psi}_\mu\gamma^\mu)\,\delta(\gamma_\mu\psi^\mu)
\,\exp\left\{i\,S\right\}\exp(-i \,m\int d^4x[
\bar{\psi}_\mu\partial^\mu\epsilon+\bar{\epsilon}\,\partial^\mu\psi_\mu]).
\eeq
Finally, integrating over the spinor fields $\epsilon$ and $\bar{\epsilon}$ yields
\beq
Z={\rm const} 
\int D\psi_\mu D\bar{\psi}_\mu D\phi\, 
\delta(\gamma_\mu\psi^\mu)
\,\delta(\bar{\psi}_\mu\gamma^\mu)\,
\delta(\partial_\mu\psi^\mu)
\,\delta(\partial^\mu\bar{\psi}_\mu)\exp\left\{i\,S\right\}.
\eeq
This is exactly the same expression as given in Eq.~(46) of
Ref.~\cite{Pascalutsa:1998pw} 
which was derived for a specific gauge-invariant model using the Dirac-Faddeev
quantization framework \cite{Dirac:1964,Faddeev:70,Senjanovic:1977vr}, see also
Ref.~\cite{Wies:2006rv} for a related work. 
Notice that the interacting theory has the same constraints as the free
one. Clearly, this is the consequence of the fact that the
gauge-invariance is only broken by the mass term in both theories, see \cite{Pascalutsa:1998pw}
for an extended discussion. 

%%%%%%%%%%%%%%%%%%%%%%%%%%%%%%%%%%%%%%%%%%%%%%%%%%%%%%%%%%%%%%%%%%%%%%%%%%%%%%%%%
\section{On-shell equivalence of interactions involving spin-$3/2$ fields}
\def\theequation{\arabic{section}.\arabic{equation}}
\setcounter{equation}{0}
\label{sec2}
We now approach our main goal and demonstrate the on-shell
equivalence of gauge-invariant and
gauge non-invariant formulations of the interacting theories with the massive
spin-$3/2$ fields 
extending the earlier work \cite{Pascalutsa:2000kd} to the case of bilinear
couplings. 

Consider the Lagrangian for interacting massive spin-$3/2$ particles given
in its general form 
\beq
\label{L:original}
{\cal L}=\bar{\psi}(i\sl{\partial}-m+V)\psi,
\eeq
where $V_{\mu \nu}$ represents an arbitrary local interaction. 
We claim that such a theory\footnote{Throughout this work we refer to the
naive Feynman rules in the case of gauge non-invariant couplings.} is
on-shell equivalent to the gauge-invariant one
described by the Lagrangian
\beq
\label{L:transformed}
{\cal L}^\prime=\bar{\psi} \bigg( i\sl{\partial}-m+i\sl{\partial}
\frac{1}{m}V\frac{1}{m}\left[1-(i\sl{\partial}+m)
\frac{1}{m}V\frac{1}{m}\right]^{-1}\hspace{-0.4cm}i\sl{\partial}\bigg)
\psi.
\eeq 
To prove this statement we rely on the equivalence-theorem~\cite{Haag:1958vt,Coleman:1969sm} 
which states that two field theories are equivalent on mass shell provided there 
exists a field transformation 
\beq
\label{equivalence}
\psi\rightarrow\psi+ P [\psi, \, \phi ]
%{\rm terms\,quadratic\,or\,higher\,order\,in\,the\,fields}, 
\eeq
which transforms one Lagrangian into another. Here, $P[\psi , \, \phi] =
\mathcal{O} (\psi^2, \, \psi \phi)$ denotes a local
polynomial of the fields $\psi$, $\phi$ and their derivatives. In the following, we will show
that the original Lagrangian in Eq.~(\ref{L:original}) can be brought into the
gauge-invariant form given in Eq.~(\ref{L:transformed}) by the following field
transformation: 
\beq
\label{redef}
\psi\rightarrow S^{-1}\left[ X - 1 \right]^{-1}  
S\,\psi\quad {\rm and}\quad\bar{\psi}\rightarrow
\bar{\psi}\,\bar{S}
\left[\bar{X} - 1 \right]^{-1}
\bar{S}^{-1},
\eeq
where the operators $X$ and $\bar X$ are defined via
\beq
\label{defX}
X=1+\left[1-\frac{1}{m}V\frac{1}{m}(i\sl{\partial}
+m)\right]^{1/2}\quad{\rm and }\quad 
\bar{X}=1+\left[1-(i\sl{\partial}
+m)\frac{1}{m}V\frac{1}{m}\right]^{1/2}.
\eeq
Further, the quantities $S$ and $\bar{S}$ have the form 
\beq
\label{defS}
S=\frac{1}{2} \left[X-\frac{1}{m}V\right] =
\frac{1}{2}X\left[1-X^{-1}\frac{1}{m}V\right]  \quad
{\rm and}\quad 
\bar{S}= \frac{1}{2}\left[\bar X - V\frac{1}{m} \right] =
\frac{1}{2}\left[1-V\frac{1}{m}\bar{X}^{-1}\right]\bar{X}.
\eeq
Notice that the field redefinition specified in Eq.~(\ref{redef}) is clearly 
of the form of Eq.~(\ref{equivalence}).

To prove the equivalence of the two Lagrangians we have to show that
\beq
\bar{S}[\bar{X}-1]^{-1}\bar{S}^{-1}(i\sl{\partial}-m+V)S^{-1}[X-1]^{-1}S \stackrel{!}{=}
i\sl{\partial}-m+i\sl{\partial}
\frac{1}{m}V\frac{1}{m}\left[1-(i\sl{\partial}+m)
\frac{1}{m}V\frac{1}{m}\right]^{-1}\hspace{-0.4cm}i\sl{\partial}.
\eeq
Using Eq.~(\ref{rel4}) from Appendix~\ref{app1} to rewrite the right-hand side
of the above equation, so that the relation to be verified becomes 
\beq
\bar{S}^{-1}(i\sl{\partial}-m+V)
S^{-1} \stackrel{!}{=} [\bar{X}-1]\bar{S}^{-1}(i\sl{\partial}-m+V)S^{-1}[X-1]+
4(i\sl{\partial}+m)-2[\bar{X}-1]\bar{S}^{-1}(i\sl{\partial}+m)
-2(i\sl{\partial}+m)S^{-1}[X-1].
\eeq
The left-hand side of this relation can be rewritten in the following way:
\beqa
\label{int0}
\bar{S}^{-1}(i\sl{\partial}-m+V) S^{-1} &=& 
\bar{S}^{-1}\big( 2\bar{S}\bar{X}^{-1}(i\sl{\partial}+m)
-2 m S   \big) S^{-1} \nn
&=& 2(\bar{X}^{-1}(i\sl{\partial}+m)
S^{-1}-\bar{S}^{-1}m) \,.
\eeqa
In the first line we used Eq.~(\ref{int0rel}) of Appendix~\ref{app1}.
The relation to prove then becomes
\beq
\label{int1}
4(i\sl{\partial}+m)-2[\bar{X}-1]\bar{S}^{-1}(i\sl{\partial}+m)
-2\bar{X}^{-1}(i\sl{\partial}+m)S^{-1}X+2\bar{S}^{-1}m X
-2\bar{X}\bar{S}^{-1}m [X-1]\stackrel{!}{=} 0.
\eeq
Using the commutation-like relation
\beq
m\left[1-X^{-1}\frac{1}{m}V\right]^{-1}=\left[1-V\frac{1}{m}
\bar{X}^{-1}\right]^{-1}\hspace{-0.4cm}m
\eeq
we can express $S^{-1}$ as follows
\beq
S^{-1}=2\left[1-X^{-1}\frac{1}{m}V\right]^{-1}\hspace{-0.4cm}X^{-1}=
2\frac{1}{m}\left[1-V\frac{1}{m}
\bar{X}^{-1}\right]^{-1}\hspace{-0.4cm}m X^{-1}=\frac{1}{m}
V\frac{1}{m}\bar{S}^{-1}m X^{-1}+2 X^{-1}.
\eeq
Let us denote the left-hand side of Eq.~(\ref{int1}), which we want to show to
be equal to zero, by $Y$.  
Inserting the above expression for $S^{-1}$ into the third term in Eq.~(\ref{int1})
we obtain: 
\beq
\label{int2}
Y = 4(i\sl{\partial}+m)-2[\bar{X}-1]\bar{S}^{-1}(i\sl{\partial}+m)
-2\bar{X}^{-1}(i\sl{\partial}+m)\frac{1}{m}V\frac{1}{m}
\bar{S}^{-1}m-4\bar{X}^{-1}(i\sl{\partial}+m)+2\bar{S}^{-1}m X
-2\bar{X}\bar{S}^{-1}m [X-1]\,.
\eeq
For a further simplification, we need yet another useful relation which reads
\beq
\label{int2a}
\bar{X}^{-1}(i\sl{\partial}+m)\frac{1}{m}V\frac{1}{m}=2-\bar{X}\,,
\eeq
and can be easily verified by multiplying both sides with $\bar{X}$ and using
the definition Eq.~(\ref{defX}).
We now apply this relation to rewrite the third
term in Eq.~(\ref{int2}) which leads to  
\beqa
\label{XVToS}
Y &=& 4(i\sl{\partial}+m)-2[\bar{X}-1]\bar{S}^{-1}(i\sl{\partial}+m)
-4\bar{X}^{-1}(i\sl{\partial}+m)
-2[\bar{X}-1]\bar{S}^{-1}m [X-2] \nn
&=& 4(i\sl{\partial}+m)-2[\bar{X}-1]\bar{S}^{-1}(i\sl{\partial}+m)
-4\bar{X}^{-1}(i\sl{\partial}+m)
+2[\bar{X}-1]\bar{S}^{-1}m X^{-1}\frac{1}{m}V\frac{1}{m}
(i\sl{\partial}+m) \nn
&=&4(i\sl{\partial}+m)-2[\bar{X}-1]\bar{S}^{-1}(i\sl{\partial}+m)
-4\bar{X}^{-1}(i\sl{\partial}+m)
+2[\bar{X}-1]\bar{S}^{-1}V\frac{1}{m}\bar{X}^{-1}
(i\sl{\partial}+m)\,.
\eeqa
Here, in the second line we again used Eq.~(\ref{int2a}) to rewrite $X - 2$. 
To end up with the third line we used the following commutation-like relation
\beq
X^{-1}\frac{1}{m}V\frac{1}{m}=\frac{1}{m}V\frac{1}{m}
\bar{X}^{-1}.
\eeq
Finally, putting together the second and the last term in the last line of
Eq.~(\ref{XVToS}) yields the desired result:
\beqa
Y &=& 4(i\sl{\partial}+m)-4[\bar{X}-1]\bar{S}^{-1}\bar{S}\bar{X}^{-1}(i\sl{\partial}+m)
-4\bar{X}^{-1}(i\sl{\partial}+m) \nn
&=& 0 \,.
\eeqa
Thus, the redefinition of the spin-$3/2$ field specified in Eq.~(\ref{redef}) in
the underlying gauge non-invariant Lagrangian $\cal{L}$ in
Eq.~(\ref{L:original}) leads to the gauge-invariant form $\cal{L}^\prime$ given in
Eq.~(\ref{L:transformed}). Notice that even if there is only a finite number of
interaction terms in the original Lagrangian, the transformed Lagrangian
always contains an infinite number of terms. 

Let us now briefly discuss possible linear couplings like the couplings of the
$\Delta$ to a nucleon fields. Once the bilinear terms are gauge-invariant
the entire Lagrangian can be transformed into
gauge-invariant form by much simpler transformation which was extensively
discussed by Pascalutsa~\cite{Pascalutsa:2000kd}. We briefly repeat the 
arguments. We start with the Lagrangian
\beq
{\cal L}=\bar{\psi}(i\sl{\partial}-m+V)\psi+\bar{\psi}{\cal O}_{\Delta N}N
+\bar{N}{\cal O}_{N \Delta}\psi+\dots,
\eeq
where $N$ and $\bar{N}$ represent nucleon fields, the operators $V$, ${\cal
  O}_{\Delta N}$, ${\cal O}_{N \Delta}$ represent interactions between 
nucleons and deltas and pions or other fields and the ellipsis represent all other
possible interactions which do not include delta degrees of freedom. After the 
transformation Eq.~(\ref{redef}) we obtain the Lagrangian
\beq
{\cal L}^\prime=\bar{\psi}(V_{\Delta\Delta}-m)\psi+\bar{\psi}\tilde{V}_{\Delta N}N
+\bar{N}\tilde{V}_{N\Delta}\psi+\dots,
\eeq
where the operators $V_{\Delta\Delta}$, $\tilde{V}_{\Delta N}$ and
$\tilde{V}_{N\Delta}$ are given as follows
\beqa
V_{\Delta\Delta}&=&i\sl{\partial}+i\sl{\partial}
\frac{1}{m}V\frac{1}{m}\left[1-(i\sl{\partial}+m)
\frac{1}{m}V\frac{1}{m}\right]^{-1}\hspace{-0.4cm}i\sl{\partial},\, \nn
\tilde{V}_{\Delta N}&=&\bar{S}
\left[\bar{X} - 1 \right]^{-1}
\bar{S}^{-1}{\cal O}_{\Delta N},\, \nn
\tilde{V}_{N \Delta}&=&{\cal O}_{N \Delta}S^{-1}\left[ X - 1 \right]^{-1}  
S~.
\eeqa
Note that the interaction $V_{\Delta\Delta}$ includes only gauge-invariant
vertices. However the interactions $\tilde{V}_{\Delta N}$ and 
$\tilde{V}_{N \Delta}$ are in general not gauge-invariant. In order to bring
them into a gauge-invariant form let us shift the delta fields like
\beq
\bar{\psi}_\mu\rightarrow\bar{\psi}_\mu+\bar{\xi}_\mu, \quad
\psi_\mu\rightarrow\psi_\mu + \xi_\mu,
\eeq
where $\bar{\xi}_\mu$ and $\xi_\mu$ are not yet specified but 
assumed to have the property not to depend on the delta fields. 
The transformed Lagrangian reads
\beq
{\cal L}^{\prime\prime}=\bar{\psi}(V_{\Delta\Delta}-m)\psi
+\bar{\xi}V_{\Delta\Delta}\psi+\bar{\psi}V_{\Delta\Delta}\xi+\bar{\psi}(\tilde{V}_{\Delta N}N-m\xi)
+(\bar{N}\tilde{V}_{N\Delta}-\bar{\xi}m)\psi+\ldots~.
\eeq  
In order to get a gauge-invariant form we now fix the fields $\xi_\mu$ and
$\bar{\xi}_\mu$ to
\beq
\bar{\xi}=\bar{N}\tilde{V}_{N \Delta}\frac{1}{m}, \quad
\xi=\frac{1}{m}\tilde{V}_{\Delta N}N~.
\eeq
With this choice the Lagrangian becomes gauge-invariant:
\beq
{\cal L}^{\prime\prime}=\bar{\psi}(V_{\Delta\Delta}-m)\psi+
\bar{\psi}V_{\Delta N}N+\bar{N}V_{N \Delta}\psi+\ldots~,
\eeq
where the gauge-invariant delta-nucleon interactions are given by
\beq
V_{\Delta N}=V_{\Delta\Delta}\frac{1}{m}\tilde{V}_{\Delta N},\quad
V_{N \Delta}=\tilde{V}_{N \Delta}\frac{1}{m}V_{\Delta\Delta}~.
\eeq

The equivalence of the two
formulations by means of the non-linear field redefinition as discussed above has important
consequences for calculations within chiral effective field theory. It implies
that S-matrix elements can be calculated from the standard effective Lagrangian
with the chiral symmetry being realized on a term-by-term basis using 
\emph{naive} Feynman rules.   

%%%%%%%%%%%%%%%%%%%%%%%%%%%%%%%%%%%%%%%%%%%%%%%%%%%%%%%%%%%%%%%%%%%%%%%%%%%%%%%%%
\section{Dependence on off-shell parameters}
\def\theequation{\arabic{section}.\arabic{equation}}
\setcounter{equation}{0}
\label{sec:offshell}

Let us briefly discuss what happens with the off-shell parameters which
account for the unphysical spin-$1/2$ degrees of freedom. The original interaction $\bar{\psi}V\psi$
can be written as a series of interactions with different off-shell parameters
$Z_i$ and $Z_i^\prime$
\beq
\bar{\psi}V\psi=\sum_i \bar{\psi}\Theta(Z_i^\prime)V_i\Theta(Z_i)\psi + {\rm h.c.}, \quad \Theta(Z_i)_{\mu\nu}=g_{\mu\nu}+Z_i\gamma_\mu\gamma_\nu.
\eeq
Multiplication of the off-shell function $\Theta(Z_i)$ with
$\frac{1}{m}i\sl{\partial}\psi$ or $\bar{\psi}i\sl{\partial}\frac{1}{m}$ from
right or left, respectively, projects out the off-shell parameters $Z_i$:
\beqa
\left[\Theta(Z_i)\frac{1}{m}i\sl{\partial}\psi\right]_\mu&=&\left[\frac{1}{m}i\sl{\partial}\psi\right]_\mu+Z_i\gamma_\mu\gamma^\lambda
\left[\frac{1}{m}\right]_{\lambda\nu}\left[i\sl{\partial}\psi\right]^\nu \nn
&=&
\left[\frac{1}{m}i\sl{\partial}\psi\right]_\mu+\frac{Z_i}{m}\frac{1}{d-1}\gamma_\mu\gamma_\nu\left[i\sl{\partial}\psi\right]^\nu\nn
&=&\left[\frac{1}{m}i\sl{\partial}\psi\right]_\mu+\frac{Z_i}{m}\frac{d-2}{d-1}\gamma_\mu\gamma^{\nu\lambda}i\partial_\lambda\psi_\nu \nn
&=&
\left[\frac{1}{m}i\sl{\partial}\psi\right]_\mu+\frac{Z_i}{m}\frac{d-2}{d-1}i\gamma_\mu(\partial^\nu-\partial_\lambda\gamma^\lambda\gamma^\nu)\psi_\nu \nn
&=& \left[\frac{1}{m}i\sl{\partial}\psi\right]_\mu.\quad
\eeqa
In the third step we used the relation
\beq
\gamma_\alpha\gamma^{\alpha\nu\lambda}=(d-2)\gamma^{\nu\lambda}.
\eeq
In the last step the field constraints
\beq
\gamma^\nu\psi_\nu=\partial^\nu\psi_\nu=0
\eeq
 have been used. Similarly one can show that
\beq
\left[\bar{\psi}i\sl{\partial}\frac{1}{m}\Theta(Z_i^\prime)\right]_\mu=\left[\bar{\psi}i\sl{\partial}\frac{1}{m}\right]_\mu
\eeq 
by using the field constraints
\beq
\bar{\psi}^\nu\gamma_\nu=\bar{\psi}^\nu\partial_\nu=0.
\eeq
Due to these relations all the off-shell parameter contributions on the most
left/right hand side of the gauge-invariant interactions
$V_{\Delta\Delta}$, $V_{\Delta N}$ and $V_{N\Delta}$ are projected
out. However, all the other off-shell parameter contributions are not zero but
can be accounted for by counterterms.
So in order for a given field theory to be independent of the off-shell
parameters counterterms are needed to absorb all the off-shell
dependence. The theory becomes automaticaly an effective field theory with
infinitely many counterterms. In this picture, the off-shell parameters play
the role of renormalization scale: the counterterms do explicitely depend on
the off-shell parameters in a way that all the calculated observables remain 
independent on them. 

In the case of the theory with the chiral Lagrangian it remains to be
clarified whether
the originally present counterterms are sufficient in order to absorb all the
off-shell parameter contributions. Note that the gauge-invariant interactions 
$V_{\Delta\Delta}$, $V_{\Delta N}$ and $V_{N\Delta}$ involve not only
covariant derivatives which appear in the chiral Lagrangian. For this
reason, it is not obvious that the chiral-invariant counterterms in the
original Lagrangian can completely compensate the off-shell dependence. 
For a related discussion we refer to the work of Ellis and Tang
\cite{Tang:1996sq,Ellis:1996bd}. Notice further that at least in the case of 
pion-nucleon scattering at leading one-loop order the off-shell parameters
can be absorbed in low-energy constants accompanying certain local interaction 
operators. This was confirmed in the explicit calculation by Fettes and 
Mei{\ss}ner \cite{Fettes:2000bb} based on the so-called small scale expansion 
\cite{Hemmert:1997ye}. 

%It may be possible that additional
%counterterms are needed to absorb the off-shell dependence which do not 
%come from chiral Lagrangian and do not respect chiral symmetry.  

%%%%%%%%%%%%%%%%%%%%%%%%%%%%%%%%%%%%%%%%%%%%%%%%%%%%%%%%%%%%%%%%%%%%%%%%%%%%%%%%%
\section{Interpretation in terms of Feynman diagrams}
\begin{figure}[tb]
\vskip 1 true cm
%  \begin{center} 
%    \epsfxsize=4.3cm
%    \epsffile{fig2.eps}
\includegraphics[width=6cm,keepaspectratio,angle=0,clip]{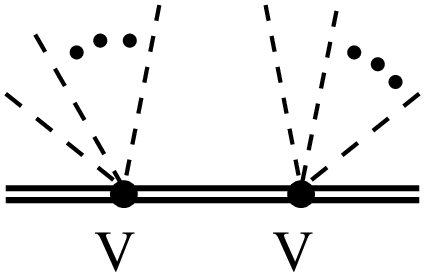}
    \caption{
         Feynman diagram to illustrate the relation between gauge
         non-invariant and gauge-invariant formulations as described in the
         text. The solid double line represents the delta, while the
         dashed lines are arbitrary external fields like pions, photons and so
         on.
\label{fig1} 
 }
%  \end{center}
\end{figure}

\def\theequation{\arabic{section}.\arabic{equation}}
\setcounter{equation}{0}
\label{sec3}
The general result proved algebraically in the previous section can also be
understood on the level of Feynman diagrams.
The underlying mechanism is even more evident in that case. To be specific,
consider the generic Feynman diagram shown in  Fig.~\ref{fig1} on mass-shell.
We now show that the manipulations discussed in the previous section simply
correspond to the  partial fraction decomposition of the
free Feynman propagator. 
The corresponding expression in operator form reads: 
\beq
i\sl{\partial}\frac{1}{m}i\,V\frac{i}{i\sl{\partial}-m}\,i\,V\frac{1}{m}i\sl{\partial}=i\sl{\partial}\frac{1}{m}i\,V\frac{1}{m}m\frac{i}{i\sl{\partial}-m}m\frac{1}{m}\,i\,V\frac{1}{m}i\sl{\partial}.
\eeq
Using the partial fraction decomposition
\beq
m\frac{1}{i\sl{\partial}-m}m=-(i\sl{\partial}+m)+
i\sl{\partial}\frac{1}{i\sl{\partial}-m}i\sl{\partial}
\eeq
we can rewrite this diagram in a gauge-invariant form:
\beq
i\sl{\partial}\frac{1}{m}i\,V\frac{i}{i\sl{\partial}-m}\,i\,V
\frac{1}{m}i\sl{\partial}=i\sl{\partial}\frac{1}{m}
i\,V\frac{1}{m}i\sl{\partial}
\frac{i}{i\sl{\partial}-m}i\sl{\partial}\frac{1}{m}i\,V
\frac{1}{m}i\sl{\partial}+
i\sl{\partial}\frac{1}{m}
i\,V\frac{1}{m}(i\sl{\partial}+m)\frac{1}{m}V\frac{1}{m}i\sl{\partial}.
\eeq
Notice that because of the appearance of the operators $\frac{1}{m}i\sl{\partial}$ and
$i\sl{\partial}\frac{1}{m}$ the first part 
includes only gauge invariant vertices which can not depend on off-shell 
parameters. The second term is just a vertex which accounts for all the 
off-shell parameters. So by partial fraction decomposition one can naturally
split the true spin-$3/2$ propagating structures which are gauge-invariant
and all non-propagating spin-$1/2$ contributions which are collected into additional
vertices.

%It is interesting to see that the original theory which has
%a lot of diseases~\cite{Velo:1969bt}\cite{Johnson:1960vt} 
%is completely equivalent on the mass shell to the gauge-invariant theory
%which seems to be free of all the inconsistencies\cite{Pascalutsa:1998pw}. 
%One can conclude that
%all the inconsistencies of the original theory like e.g. acausality of
%classical spin-$3/2$ propagator in the presence of a background field show
%up only off-shell.
%Now we give a general relation between the gauge-invariant and original
%propagator in the presence of background interaction $V$.

%{\bf Lemma 2}:
%%%%%%%%%%%%%%%%%%%%%%%%%%%%%%%%%%%%%%%%%%%%%%%%%%%%%%%%%%%%%%%%%%%%%%%%%%%%%%%%%
\section{Summary and conclusions}
\def\theequation{\arabic{section}.\arabic{equation}}
\label{sec:summary}

In this paper, we have explicitely constructed a non-linear field transfromation
that brings the interaction Lagrangian of spin-$3/2$ fields as given in 
Eq.~(\ref{L:original}) in a form that is invariant under gauge transformation
Eq.~(\ref{gaugetrafo}). This  proves a conjecture made in
Ref.~\cite{Pascalutsa:2000kd} where it was shown how a gauge non-invariant
Lagrangian with linear couplings in the spin-$3/2$ fields can be brought into
a gauge-invariant form via a suitably chosen field redefinition. We have also
given a diagrammatic explanation of our main result, which allows to split
the truly propagating spin-$3/2$ components from the non-propagating
spin-$1/2$ contributions that are subsumed in local operators existing in the
most general Lagrangian of deltas coupled to nucleons, pions and external
fields.

%%%%%%%%%%%%%%%%%%%%%%%%%%%%%%%%%%%%%%%%%%%%%%%%%%%%%%%%%%%%%%%%%%%%%%%%%%%%%%%%%
\section*{Acknowledgments}

We are grateful to Jambul Gegelia and Vladimir Pascalutsa for useful comments 
on the manuscript. The work of E.E. and H.K. was supported in parts by funds provided from the 
Helmholtz Association to the young investigator group  
``Few-Nucleon Systems in Chiral Effective Field Theory'' (grant  VH-NG-222)
and through the virtual institute ``Spin and strong QCD'' (grant VH-VI-231). 
This work was further supported by the DFG (SFB/TR 16 ``Subnuclear Structure
of Matter'') and by the EU Integrated Infrastructure Initiative Hadron
Physics Project under contract number RII3-CT-2004-506078.

%%%%%%%%%%%%%%%%%%%%%%%%%%%%%%%%%%%%%%%%%%%%%%%%%%%%%%%%%%%%%%%%%%%%%%%%%%%%%%%%%
\appendix

\def\theequation{\Alph{section}.\arabic{equation}}
\setcounter{equation}{0}
\section{Some useful operator relations}
\label{app1}

In this appendix we list a number of helpful relations involving the operators $X$,
$\bar X$, $S$ and $\bar S$, see Eqs.~(\ref{defX},\ref{defS}),
\beq
\label{rel1}
2\bar{S}[\bar{X}-1]^{-1}-1=\left(1-V\frac{1}{m}\right)[\bar{X}-1]^{-1},
\quad
[X-1]^{-1}2S-1=[X-1]^{-1}\left(1-\frac{1}{m}V\right),
\eeq
\beq
\label{rel2}
[\bar{X}-1]^{-2}m=[\bar{X}-1]^{-2}(i\sl{\partial}+m)\frac{1}{m}V
+m,
\eeq
as well as commutation-like relation
\beq
\label{rel3}
(i\sl{\partial}+m)X=\bar{X}(i\sl{\partial}+m).
\eeq
We now use the above equations to prove another important operator relation
\beq
\label{rel4}
i\sl{\partial}
\frac{1}{m}V\frac{1}{m}\left[
\bar{X}-1
\right]^{-2} i\sl{\partial}
=[2\,\bar{S}[\bar{X}-1]^{-1}-1](i\sl{\partial}+m)[[X-1]^{-1}2\,S-1]-i\sl{\partial}-m+V \,,
\eeq
used in section  \ref{sec2} to show the on-shell equivalence of the gauge-invariant
and the gauge non-invariant bilinear interactions of the massive spin-$3/2$ field. 
A straightforward calculation yields: 
\beqa
i\sl{\partial}
\frac{1}{m}V\frac{1}{m}[\bar{X}-1]^{-2}i\sl{\partial}&=&
(i\sl{\partial}+m)\frac{1}{m}V\frac{1}{m}
[\bar{X}-1]^{-2}i\sl{\partial}-V\frac{1}{m}[\bar{X}-1]^{-2}i\sl{\partial} \nn
&=&[1-[\bar{X}-1]^2][\bar{X}-1]^{-2}i\sl{\partial}-
V\frac{1}{m}[\bar{X}-1]^{-2}i\sl{\partial} \nn
&=&
\left(1-V\frac{1}{m}\right)[\bar{X}-1]^{-2}i\sl{\partial}-i\sl{\partial} \nn
&=& \left(1-V\frac{1}{m}\right)[\bar{X}-1]^{-2}(i\sl{\partial}+m)
- \left(1-V\frac{1}{m}\right)[\bar{X}-1]^{-2}m-
i\sl{\partial} \nn
&=& \left(1-V\frac{1}{m}\right)[\bar{X}-1]^{-2}
(i\sl{\partial}+m)-i\sl{\partial}
-\left(1-V\frac{1}{m}\right)
\left([\bar{X}-1]^{-2}(i\sl{\partial}+m)\frac{1}{m}V+m\right)
\nn
&=&\left(1-V\frac{1}{m}\right)[\bar{X}-1]^{-1}(i\sl{\partial}+m)
[X-1]^{-1}\left(1-\frac{1}{m}V\right)-i\sl{\partial}-m+V.
\eeqa
Here, in the second, fifth and sixth lines we used Eqs.~(\ref{defX}), 
(\ref{rel2}) and (\ref{rel3}), respectively. Inserting Eq.~(\ref{rel1})
into the last line of the above equation we complete the proof of 
Eq.~(\ref{rel4}).  

At the end of this appendix we would like to prove an additional relation
used in the first line of Eq.~(\ref{int0})
\beq
\label{int0rel}
i\sl{\partial}-m+V=2\bar{S}\bar{X}^{-1}(i\sl{\partial}+m)
-2 m S~.
\eeq
To derive this relation we use the commutation-like relation
\beq
(i\sl{\partial}+m)X^{-1}=\bar{X}^{-1}(i\sl{\partial}+m)
\eeq
as well as the ``conjugate'' of Eq.~(\ref{int2a})
\beq
\label{int2aconj}
\frac{1}{m}V\frac{1}{m}(i\sl{\partial}+m)X^{-1}=2-X\,~.
\eeq
With these relations the derivation of the first
line becomes obvious: 
\beqa
i\sl{\partial}-m+V&=&i\sl{\partial}+m-2m+V=i\sl{\partial}+m-m(2-X)-2m S\nn
&=&
i\sl{\partial}+m-V\frac{1}{m}(i\sl{\partial}+m)X^{-1}-2m S
=i\sl{\partial}+m-V\frac{1}{m}\bar{X}^{-1}(i\sl{\partial}+m)-2m S\nn
&=&
(1-V\frac{1}{m}\bar{X}^{-1})(i\sl{\partial}+m)-2m S=
2\bar{S}\bar{X}^{-1}(i\sl{\partial}+m)
-2 m S~.
\eeqa
%%%%%%%%%%%%%%%%%%%%%%%%%%%%%%%%%%%%%%%%%%%%%%%%%%%%%%%%%%%%%%%%%%%%%%%%%%%%%%%%%

%\setlength{\bibsep}{0.2em}
%\bibliographystyle{h-physrev3}
%\bibliography{/home/epelbaum/refs_h-elsevier3}
%\end{document}

\end{document}